\documentclass[11pt,a4paper]{article}
\pdfoutput = 1
\usepackage{jheppub}
\usepackage{color}
\usepackage{amsmath,bm}
\usepackage{graphicx}

\newcommand{\as}{\alpha_{\mathrm{s}}}
\newcommand{\LA}{\mathrm{A}}

\newcommand{\LF}{\mathrm{F}}

\newcommand{\LR}{\mathrm{R}}

\newcommand{\LT}{\mathrm{T}}
\newcommand{\LZ}{\mathrm{Z}}

\newcommand{\Lb}{\mathrm{b}}
\newcommand{\Lc}{\mathrm{c}}

\newcommand{\Lf}{\mathrm{f}}
\newcommand{\Lg}{\mathrm{g}}

\definecolor{red}{rgb}{1,0,0}

\newcommand{\MSbar}{\overline {\text{MS}}}

\newcommand{\GeV}{\ \mathrm{GeV}}
\newcommand{\TeV}{\ \mathrm{TeV}}

\def\<>#1{\big\langle{#1}\big\rangle}
\def\[]#1{\big[{#1}\big]}

\newbox\charbox
\newbox\slabox
\def\s#1{{      
        \setbox\charbox=\hbox{$#1$}
        \setbox\slabox=\hbox{$/$}
        \dimen\charbox=\ht\slabox
        \advance\dimen\charbox by -\dp\slabox
        \advance\dimen\charbox by -\ht\charbox
        \advance\dimen\charbox by \dp\charbox
        \divide\dimen\charbox by 2
        \raise-\dimen\charbox\hbox to \wd\charbox{\hss/\hss}
        \llap{$#1$}
}}

\title{A parton shower based on factorization of the quantum density matrix}

\author[a]{Zolt\'an Nagy}

\author[b]{and Davison E.\ Soper}

\affiliation[a]{
DESY\\
Notkestrasse 85\\
22607 Hamburg, Germany
}

\affiliation[b]{
Institute of Theoretical Science\\
University of Oregon\\
Eugene, OR  97403-5203, USA
}

\emailAdd{Zoltan.Nagy@desy.de}
\emailAdd{soper@uoregon.edu}

\abstract{
We present first results from a new parton shower event generator, \textsc{Deductor}. Anticipating a need for an improved treatment of parton color and spin, the structure of the generator is based on the quantum density matrix in color and spin space. So far, \textsc{Deductor} implements only a standard spin-averaged treatment of spin in parton splittings. Although \textsc{Deductor} implements an improved treatment of color, in this paper we present results in the standard leading color approximation so that we can compare to the generator \textsc{Pythia}. The algorithms used incorporate a virtuality based shower ordering parameter and massive initial state bottom and charm quarks.
}

\preprint{DESY 13-241}

\begin{document}
\maketitle

\section{Introduction}
\label{sec:intro}

Parton shower Monte Carlo event generators, such as \textsc{Herwig}  \cite{Herwig}, \textsc{Pythia} \cite{Pythia}, and \textsc{Sherpa} \cite{Sherpa}, have proven to be enormously useful since the development of the main ideas in the 1980s \cite{EarlyPythia, Gottschalk, angleorder}. These computer programs perform calculations of cross sections according to an approximation to the standard model or its possible extensions.  In a parton shower, one can think of the shower developing with decreasing values of a parameter that, in \textsc{Pythia} and \textsc{Sherpa}, is a measure of the hardness of interactions: smaller hardness corresponds to a larger scale of space-time separations.\footnote{\textsc{Herwig} rearranges the ordering of splittings in its shower so that larger angle splittings come first.} At the hard interaction, there are just a few partons (typically quarks and gluons). Then, as the hardness decreases, these partons split, making more partons in a parton shower.\footnote{Thus, with respect to initial state partons, the shower evolution starts from the hard interaction and moves backward in time to softer initial state interactions. With respect to final state partons, shower evolution moves forward in time.}

Because of the great success of these parton shower programs, it is worthwhile to investigate possible improvements. A few years ago, we proposed a theoretical structure for the dynamics of a parton shower that generalizes the structure of current algorithms and allows improvement over certain approximations used currently \cite{NSI}. The development of the simplest shower is based on probabilities for parton splittings. Thus the dynamics of a simple parton shower is described by classical statistical mechanics. However, the partons of quantum field theory carry color and spin, which are quantum variables. For this reason, the theoretical structure of ref.~\cite{NSI} describes the color and spin evolution of the partons using the language of quantum statistical mechanics.\footnote{Nevertheless, it suffices to use classical statistical mechanics to describe the momenta and flavors of the partons. Within the approximations of a parton shower, interference between different momentum or flavor states is not important.} The approach of ref.~\cite{NSI} contrasts with standard parton shower dynamics, in which one averages over spins and treats color using what is known as the leading color approximation.

The theoretical structure of ref.~\cite{NSI} consists of integral equations that specify the dynamics of the quantum density operator that represents the state of the shower at the current value of the hardness variable. It is not a computer program. Indeed, it is not simple to design a computer program that implements these integral equations in a practical fashion. However, one can make progress.

The first step is to construct a parton shower algorithm in a style that fits with the general theoretical structure of ref.~\cite{NSI}, but uses the leading color approximation and averages over spins. We described the core of the needed algorithms in ref.~\cite{NSII}. In this paper, we present some first results from an implementation of this algorithm in a computer program, called \textsc{Deductor}\footnote{``Deductor'' is Latin for ``guide'' or ``teacher'' in the spirit of ``Pythia'' and ``Sherpa.''} \cite{DeductorCode}.

The second step is to go beyond the leading color (LC) approximation. In ref.~\cite{NScolor}, we defined an approximation, the leading color plus (LC+) approximation, that goes beyond the LC approximation. The LC+ approximation is exact for collinear splittings and for collinear$\times$soft splittings, but still approximate for wide angle soft splittings. The LC+ approximation is implemented in \textsc{Deductor}. One can also go further order by order in a perturbatvie expasion around the LC+ approximation, although this possibility is not yet implemented in \textsc{Deductor}. In this paper we examine only results at the LC level, saving comparisons between LC+ and LC results for a later publication.

The third step is to restore quantum interference of spin amplitudes. In ref.~\cite{NSspin}, we defined an algorithm for doing that. We have not yet implemented this algorithm in \textsc{Deductor}.

\section{Description of the program}
\label{sec:description}

Based on what we have written in the Introduction, it would seem that we have simply cloned \textsc{Pythia}, which is based on similar approximations. Actually, in one sense, we have done less. We have incorporated neither a model for hadronization nor a model for the underlying event that comes with a hard scattering. Our main aim is to investigate the approximations in a parton shower algorithm and for that purpose we need only the parton shower. However, for realistic comparisons with data, one certainly needs hadronization. We anticipate linking to an external program for this purpose. We also anticipate providing a way to generate an underlying event.

The structure of \textsc{Deductor} is similar to that of \textsc{Pythia} or \textsc{Sherpa}. All three of these start at the hardest interaction and evolve towards softer interactions. All three account approximately for interference effects in the emission of soft gluons: the soft gluons are emitted from color dipoles.\footnote{This is not quite true in \textsc{Pythia}, but it is true for the final state shower \cite{SjostrandSkands}.} But in other respects, the programs are not the same. Most importantly, \textsc{Deductor} is designed to facilitate more advanced treatments of color and spin. There are some other differences even at the spin-averaged, leading color level. We sketch these differences below. In two cases, the differences require much more than a sketch and are not contained in our earlier papers \cite{NSI, NSII, NScolor, NSspin}, so we devote separate papers \cite{NSShowerTime, NSpartons} to them.

\subsection{Splitting functions}

The splitting functions that we use are not simply the DGLAP splitting functions with some cuts applied. Rather, we use the splitting functions from ref.~\cite{NSII}, which are the splitting functions of ref.~\cite{NSI} averaged over spins. For diagrams that do not involve interference, these are based very directly on the relevant Feynman diagrams with a projection onto physical polarizations for the off-shell parton. The idea is that if one makes minimal approximations, one may get closer to the exact amplitudes when successive splitting vertices are not strongly ordered in the hardness parameter.

\subsection{Momentum conservation}

In \textsc{Deductor}, as in other parton showering programs, the daughter particles in splittings are approximated as being on shell.  Evidently, when parton splitting is iterated, this approximation is not consistent with momentum conservation. Thus we need to take a small amount of momentum from elsewhere in the event and supply it to the previously on-shell mother parton. In \textsc{Deductor}, we take the needed momentum from all of the other partons in the event by applying a small Lorentz transformation to them.\footnote{For final state splittings, the Lorentz transformation is given in ref.~\cite{NSI}; for initial state splittings, we use a revised version given in refs.~\cite{NSShowerTime, NSDrellYan}.} Thus each parton is disturbed only slightly. In \textsc{Pythia} or \textsc{Sherpa} the needed momentum in many cases comes from a single parton. 

In initial state splittings in \textsc{Deductor}, there is also a Lorentz transform to bring the newly created initial state parton to zero transverse momentum.

\subsection{Shower ordering variable}

The splitting vertices in \textsc{Deductor} are ordered in decreasing values of a parameter $\Lambda^2$. As in \textsc{Pythia} or \textsc{Sherpa} (but not \textsc{Herwig}), the ordering is from hardest to softest. The ordering variable that we choose for the splitting of a parton with momentum $p_i$ is
\begin{equation}
\begin{split}
\label{eq:showertime}
\Lambda_i^2 ={}& \frac{p_i^2 - m_i^2}{2p_i\cdot Q_0} 
\,Q_0^2
\hskip 1.4 cm {\rm final\ state\ parton}
\;,
\\
\Lambda_i^2 ={}& \frac{|p_i^2 - m_i^2|}{2\eta_i\,p_\LA\cdot Q_0} 
\,Q_0^2
\hskip 1cm {\rm initial\ state\ parton}
\;.
\end{split}
\end{equation}
Here $Q_0$ is the total momentum of the final state partons created in the hard process that initiates the shower. For an initial state parton from a hadron with momentum $p_\LA$ (approximated as lightlike), $\eta_i$ is the momentum fraction of the parton. Thus the ordering variable is proportional to the virtuality $|p_i^2 - m_i^2|$ of the splitting divided by the energy of the mother parton $i$ as measured in the $\vec Q_0 = 0$ frame. In \textsc{Pythia} and \textsc{Sherpa}, the ordering variable is $k_\LT^2$, the squared transverse momentum of either of the daughter partons relative to the mother parton direction. Our choice is dictated by factorization: we want the relatively soft interaction of the current splitting to factor from the harder interactions of prior splittings on a graph by graph basis in a physical gauge. The reasoning behind this choice takes some explanation, so we devote a separate paper \cite{NSShowerTime} to it. 

In the case of initial state splittings, $\Lambda^2$ ordering allows a wider phase space for splittings than is available with other shower ordering choices. We examine this feature in ref.~\cite{NSShowerTime}.

\subsection{Parton masses}

\textsc{Deductor} uses the physical, non-zero, values of the charm and bottom quark masses, $m_\Lc$ and $m_\Lb$, throughout, both when the quarks are final state partons and when they are initial state partons. In contrast, \textsc{Pythia}, \textsc{Herwig}, and \textsc{Sherpa} set the masses of initial state partons to zero. The difference in these approaches can matter, for instance, in the case of a hard process that involves an initial state bottom quark. At an hard interaction scale $\sqrt{Q_0^2}$ of hundreds of GeV, $m_\Lb$ is negligible. However, as the shower proceeds to softer interactions, $m_\Lb$ is no longer negligible.  Since the bottom quark must have come from a splitting $\Lg \to \Lb + \bar\Lb$ at a lower scale, one may ask a parton shower event generator to tell us the distribution in rapidity and transverse momentum of the $\bar\Lb$. For this question, it matters that $m_\Lb \ne 0$.

There is a possible objection to letting initial state quark masses be non-zero: with non-zero quark masses and two incoming quarks, the factorization of the cross section into a hard scattering function times parton distribution functions fails \cite{Doria:1980ak}. There are infrared sensitive, non-factorizing contributions of order $m_q^2/E_0^2$, where $m_q$ can be $m_\Lb$ or $m_\Lc$ and $E_0$ is the energy of one of the quarks entering the hard interaction, viewed in the c.m.\ frame of the collision. We can understand this in the spirit of the arguments of ref.~\cite{CSS} by noting that the velocity $\beta$ of an initial state quark is given by $\beta^2 = 1 - m_q^2/E^2$, where $E$ is the quark energy. If $\beta < 1$, the classical world lines of the two quarks can be causally connected to each other, so that the quarks can change each other's color and thus change the probability of the hard interaction. However, we are happy to neglect $m_q^2$ compared to $E_0^2$. What we do {\em not} want to do is to neglect $m_q^2$ compared to $p_\LT^2$, where $p_\LT$ is the transverse momentum of a quark $q$ in an initial state $\Lg \to q + \bar q$ splitting. When such a splitting occurs, $1-\beta$ of the quark is even smaller than it was at the hard interaction because the quark has more energy. Thus the $\Lg \to q + \bar q$ splitting is very much out of causal communication with the partons in the hadron approaching from the other direction. Based on this physical argument, we expect that there is not a problem in treating the b or c quark as massive for the purpose of working out its initial state splittings.

\subsection{Evolution of the parton distribution functions}
\label{sec:partons}

\textsc{Deductor} uses the standard method \cite{EarlyPythia, Gottschalk} of generating initial state splittings: the probability for a splitting is proportional to a splitting function and to a ratio of parton distribution functions. In the numerator is the parton distribution function for the new initial state parton while in the denominator is the parton distribution function for the old initial state parton \cite{NSI}. The parton distribution functions, of course, obey their own evolution equation. For the use of parton distribution functions within shower evolution to be consistent, the kernel of the evolution equation for the parton distributions needs to be compatible with the splitting functions in the parton shower (see below).

The shower splitting functions that involve a massive quark depend on the quark mass in a nontrivial way. Unfortunately, the standard first order DGLAP splitting functions \cite{DGLAP} that give the evolution of $\MSbar$ parton distributions \cite{CSpartons} do not involve the quark masses except in the form of boundary conditions that tell how to go from five flavors to four flavors and then to three flavors. For this reason, \textsc{Deductor} uses non-$\MSbar$ parton distribution functions that obey an evolution equation in which the kernel has explicit dependence on the quark masses. 

Evidently, the question of mass dependence in the evolution of the parton distribution functions is not trivial. We therefore explain the issues in a separate paper \cite{NSpartons}. We first argue that the parton evolution kernels and the shower splitting functions need to be compatible and derive what the compatibility condition is (given the chosen shower ordering variable). This enables us to derive the mass-dependent parton evolution kernels.

One could consider deriving shower parton distribution functions with quark masses by fitting data. Such a project is outside the scope of this paper and ref.~\cite{NSpartons}.
Instead, we use a set of parton distribution functions fit to data using the methods of the \textsc{HeraFitter} group \cite{HeraFitter1, HeraFitter2} and kindly provided to us by that group. The fit uses lowest order perturbation theory and leading order $\MSbar$ evolution with $\as(M_\LZ) = 0.126$. We take the parametrization of this set of parton distributions at the starting scale $Q_{\rm fit} = 1 \GeV$ and apply the mass dependent evolution equations of ref.~\cite{NSpartons} to this starting set. The starting parametrization is available with the \textsc{Deductor} code \cite{DeductorCode}.

In ref.~\cite{NSpartons}, we investigate numerically how the shower parton distributions defined there differ from the $\MSbar$ parton distributions. At a factorization scale below the charm quark mass, they are by definition identical. We find that at high scales there are differences, but these differences are within the uncertainty associated with working only at the lowest order in $\as$. However, the parton distributions for heavy quarks in the shower scheme differ substantially from the corresponding $\MSbar$ distributions when we look at a factorization scale not far from the heavy quark mass.

\section{Some comparisons to \textsc{Pythia}}

In this section, we compare some results from \textsc{Deductor} to the equivalent results from \textsc{Pythia}, version 8.176 \cite{Pythia}. Our main purpose is to check that \textsc{Deductor} is producing results close to those of \textsc{Pythia} in cases for which we do not expect the physics differences between the two programs to matter greatly. In some cases, we expect to see some differences. It would certainly be of interest to compare \textsc{Deductor} also with \textsc{Herwig} and \textsc{Sherpa}. However, such a comparison is beyond the scope of this paper.

\subsection{Settings}
\label{sec:settings}

In \textsc{Deductor} in the comparisons that follow, we use parton distribution functions as described in section~\ref{sec:partons}. We evaluate $\as$ in parton splittings at a scale $\lambda_\LR k_\LT^2$ where $k_\LT$ is the transverse momentum in the splitting and where
\begin{equation}
\label{eq:lambdaR}
\lambda_\LR = \exp\left(- [ C_\LA(67 - 3\pi^2)- 10\, n_\Lf]/[3\, (33 - 2\,n_\Lf)]\right) \approx 0.4
\;.
\end{equation}
This helps to get the large log summation in a parton shower right \cite{NSDrellYan, CataniMCsummation}. We choose $\as(M_\LZ^2) = 0.118$  in \textsc{Deductor}. Then $\as(\lambda_\LR M_\LZ^2) \approx 0.126$, matching the $\as(M_\LZ^2)$ value that was used in fitting the parton distributions at lowest order, without $\lambda_\LR$.  We end the \textsc{Deductor} shower by vetoing parton splittings when $k_\LT^2 < 1 \GeV^2$. There is no hadronization stage, nor is there an underlying event. For cross sections involving jet production, we use $2 \to 2$ parton scattering with renormalization and factorization scales $\mu_\LR^2 = \mu_\LF^2 = (p_\LT/2)^2$.

In the comparisons that follow, in \textsc{Pythia}, we use MSTW 2008 LO parton distributions \cite{MSTW} and take the default choices for $\as$. We turn off hadronization and multiple parton interactions in \textsc{Pythia}, so that we examine only the parton shower created by the hard interaction. We set the parameter {\tt TimeShower:pTmin} to 1 GeV, so that final state showering ends at $k_\LT^2 = \GeV^2$, matching our choice in \textsc{Deductor}. The initial state shower in \textsc{Pythia} is controlled by two parameters, {\tt SpaceShower:pT0Ref} with default value 2.0 GeV and {\tt SpaceShower:pTmin} with default value 0.5 GeV. The first of these parameters provides a soft lower cutoff on the $k_\LT$ of initial state emissions, while the second provides a hard lower cutoff. If the \textsc{Pythia} initial state shower were a dipole shower, then setting {\tt SpaceShower:pT0Ref} to 0 and {\tt SpaceShower:pTmin} to 1.0 GeV would make the \textsc{Pythia} initial state shower most like the \textsc{Deductor} initial state shower  with $k^{\rm min}_\LT = 1.0 \GeV$. However, the \textsc{Pythia} initial state shower is not a dipole shower. The default choices for {\tt SpaceShower:pT0Ref} and {\tt SpaceShower:pTmin} result in what appears to be a sensible amount of initial state radiation, as we will see in section \ref{sec:DrellYan}. For this reason, we leave these two \textsc{Pythia} parameters at their default values. \textsc{Pythia} also applies a distribution of primordial transverse momentum that the initial state partons are assumed to have at the soft end of the shower. This feature is not implemented in \textsc{Deductor}, so we turn it off in \textsc{Pythia} using {\tt BeamRemnants:primordialKT = off}. 

Where we examine jets, we use the $k_\LT$ algorithm \cite{KTCatani, KTEllis} and we find the jets with the help of \textsc{fastjet} \cite{fastjet}.

\subsection{One jet inclusive cross section}
\label{sec:onejetincl}

\begin{figure}
\centerline{\includegraphics[width = 8 cm]{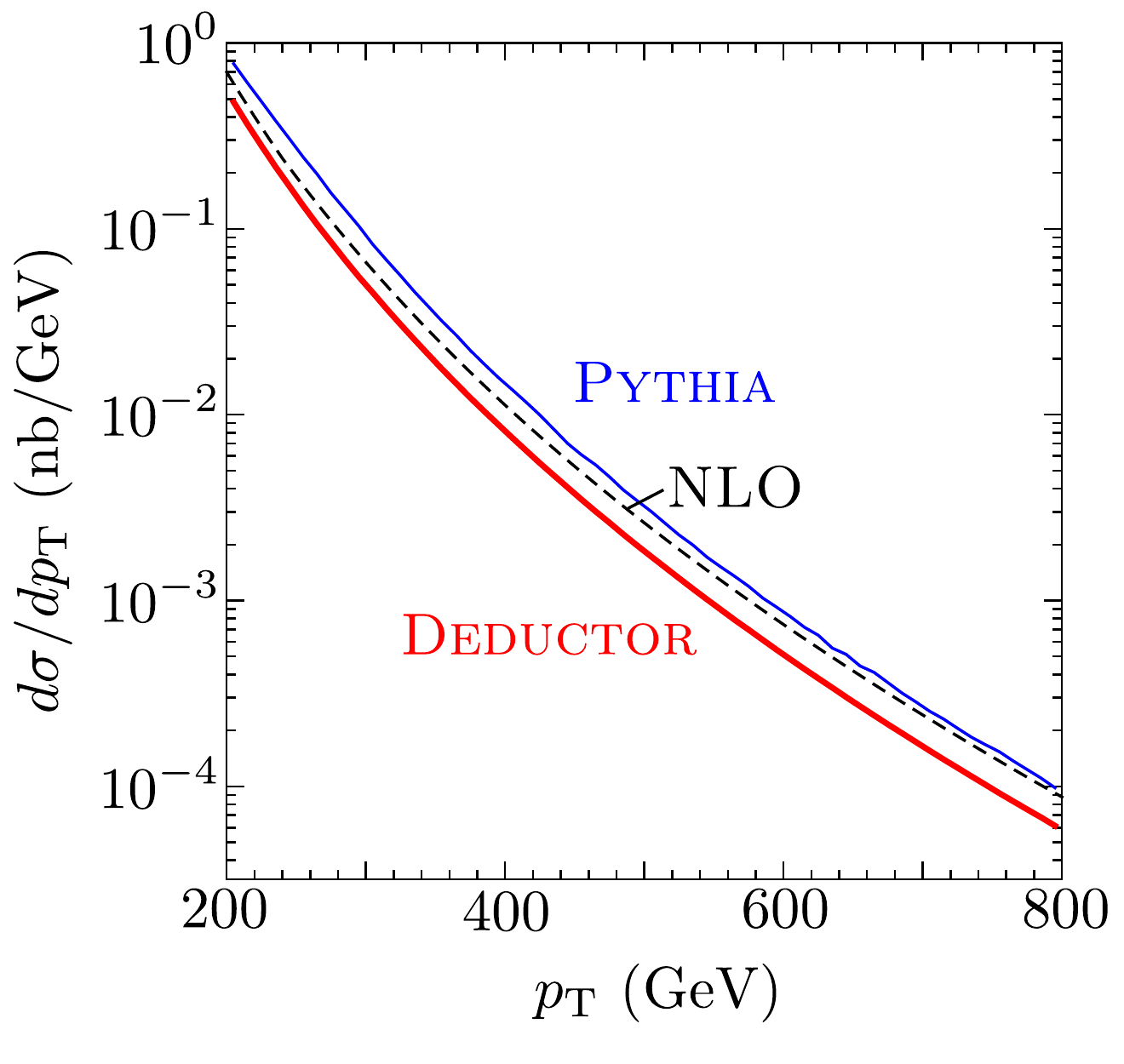}}
\caption{The one jet inclusive cross section $d\sigma/dp_\LT$ for $|y| < 2$ using the $k_\LT$ algorithm with $R = 0.4$ calculated with \textsc{Deductor} and \textsc{Pythia}. The dashed curve shows a next-to-leading order perturbative calculation.
}
\label{fig:onejet}
\end{figure}

We begin with a calculation of the one jet inclusive cross section $d\sigma/dp_\LT$ for jets in the rapidity range $|y| < 2$ in proton-proton collisions with $\sqrt s = 8 \TeV$. We use the $k_\LT$ algorithm with $R = 0.4$. We expect to get the same cross section with \textsc{Deductor} as with \textsc{Pythia} to within the accuracy of a leading order calculation, about a factor two. We see in figure~\ref{fig:onejet} that this expectation is fulfilled. We also show a perturbative next-to-leading order calculation \cite{EKS} using a scale choice $\mu_\LR^2 = \mu_\LF^2 = (p_T/2)^2$ and CT10W partons \cite{CT10}. The NLO result lies between the two parton shower results.

\subsection{Dijet angular decorrelation}

\begin{figure}
\centerline{\includegraphics[width = 8 cm]{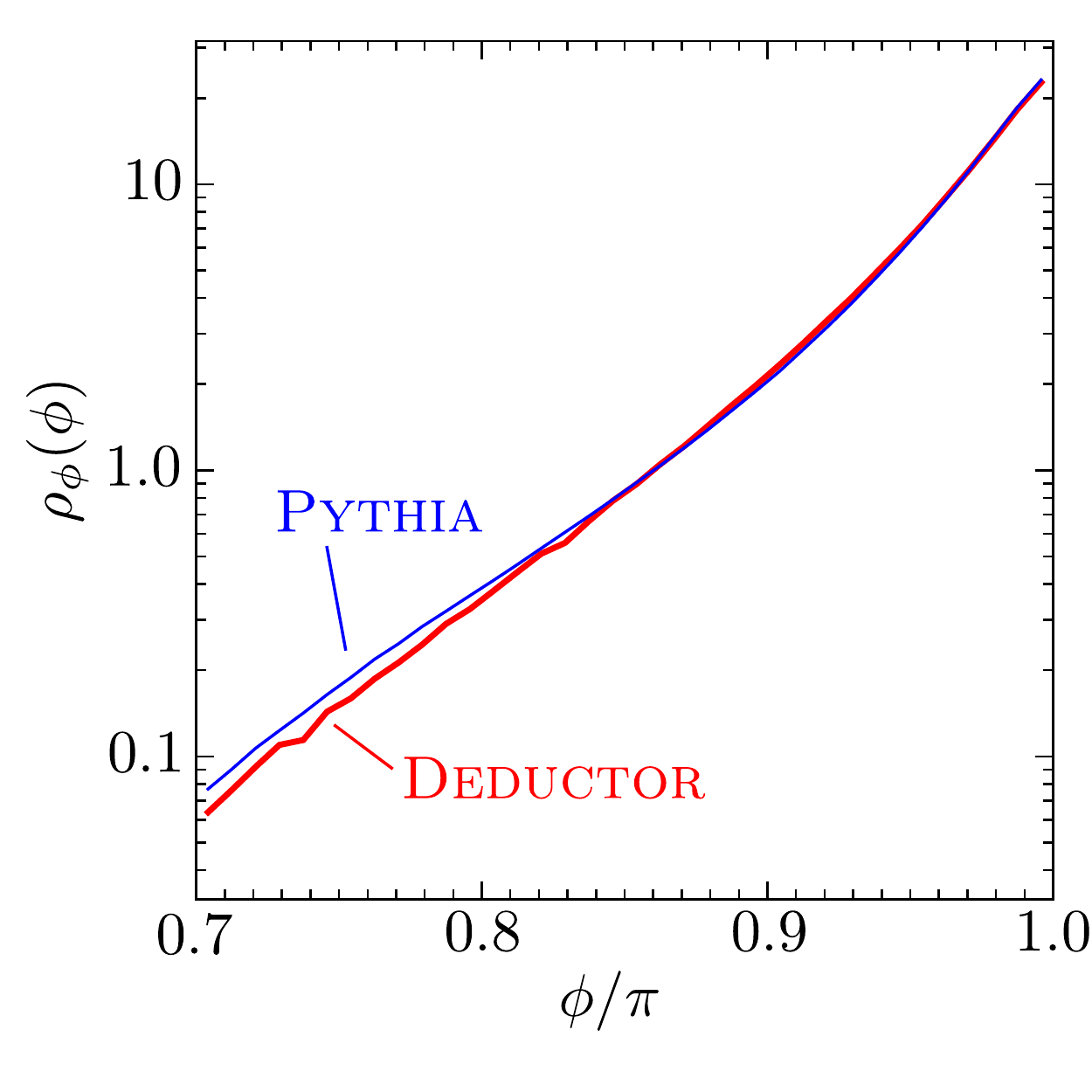}}

\caption{The angular decorrelation distribution, $\rho_\phi(\phi)$ for the two leading jets in an event with $P_\LT > 200 \GeV$ and $|y| < 2$, calculated with \textsc{Deductor} and \textsc{Pythia}.
}
\label{fig:jetangle}
\end{figure}

The dijet angular decorrelation distribution is somewhat more subtle. We consider proton-proton collisions with $\sqrt s = 8 \TeV$. We use the $k_\LT$ algorithm with $R = 0.4$ to define jets and create a sample of events having at least two jets, each with $p_\LT > 200 \GeV$ and $|y| < 2$.  Call the cross section for this $\sigma$. Let the azimuthal angle separation between the two jets with the highest $p_\LT$ be $\phi$. Then we measure the distribution $\rho_\phi(\phi) = (1/\sigma)\, d\sigma/d\phi$ for this event sample. The normalization is $\int\!d\phi\ \rho_\phi(\phi) = 1$. For an ideal two jet event, $\phi = \pi$. Emission of a soft gluon makes $\phi$ slightly less than $\pi$. Emission of a hard gluon that is not collinear with one of the two leading partons makes $\phi$ substantially less than $\pi$. For an ideal three jet event, $\phi$ can be as small as $2\pi/3$. Thus this angular distribution in the region $\phi > 2\pi/3$ but $\pi - \phi$ not too small should be reliably predicted by fixed order perturbative QCD, while a parton shower should give a good account of it for $\pi - \phi \ll 1$. For these reasons, measurement of $\rho_\phi(\phi)$ provides a good experimental test of QCD \cite{D0Angles, CMSAngles, AtlasAngles}.  In figure~\ref{fig:jetangle}, we compare the results of \textsc{Deductor} and \textsc{Pythia} for the dijet angular decorrelation. We see that the two results are very close to each other.

\subsection{Number of partons in a jet}
\label{sec:Ninjet}

\begin{figure}
\centerline{\includegraphics[width = 8 cm]{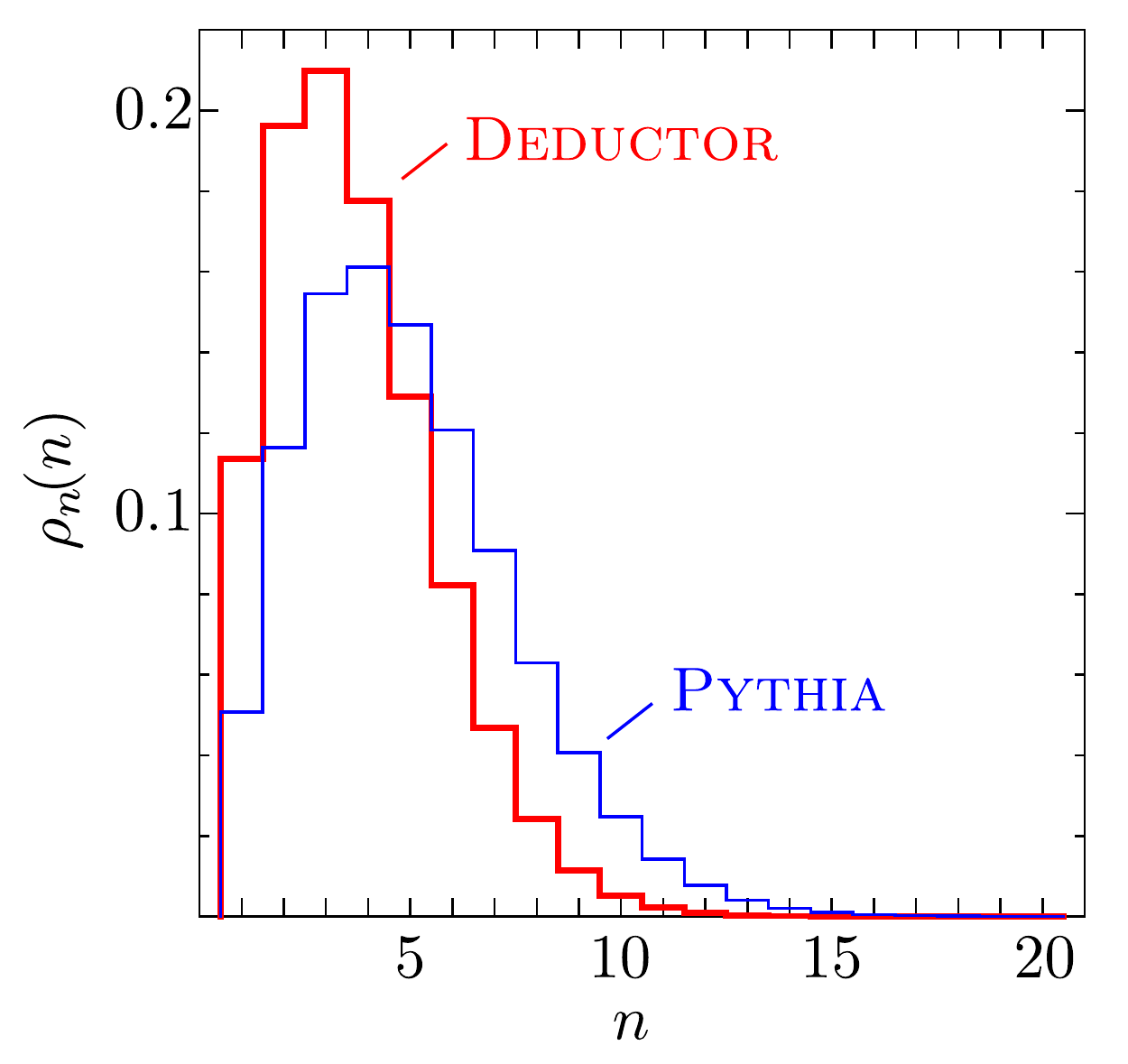}}

\caption{The distribution $\rho_n(n)$ of the number of partons in a \textsc{Pythia} jet with $P_\LT > 200 \GeV$ and $|y| < 2$ compared with the same distribution for a \textsc{Deductor} jet. The jets are constructed using the $k_\LT$ algorithm with $R = 0.4$.
}
\label{fig:Ninjet}
\end{figure}

In section \ref{sec:onejetincl} we looked at the one jet inclusive cross section. Now, we look inside these jets. We analyze a sample of jets with $P_\LT > 200 \GeV$ and $|y| < 2$. We examine the distribution $\rho_n(n)$ of the number $n$ of partons in a jet in this sample for events simulated by \textsc{Pythia} (with no hadronization or underlying event) and for events simulated by \textsc{Deductor}. The distribution is normalized to $\sum_n \rho_n(n) = 1$. Evidently, the number of partons in a jet is not a physical observable, but it of interest here because it is sensitive to the parton showering algorithms. As explained in the introduction to this section, we adjust parameters of \textsc{Pythia} to make the results as comparable as we can between \textsc{Pythia} and \textsc{Deductor}. In figure~\ref{fig:Ninjet}, we compare the results of \textsc{Deductor} and \textsc{Pythia} for the distribution of the number of partons in a jet. We see that the distributions are similar but with a peak in the distribution for \textsc{Deductor} at $n = 3$ and for \textsc{Pythia} at $n = 4$.

\subsection{$z$-distribution of partons in a jet}
\label{sec:Zinjet}

\begin{figure}

\centerline{\includegraphics[width = 8 cm]{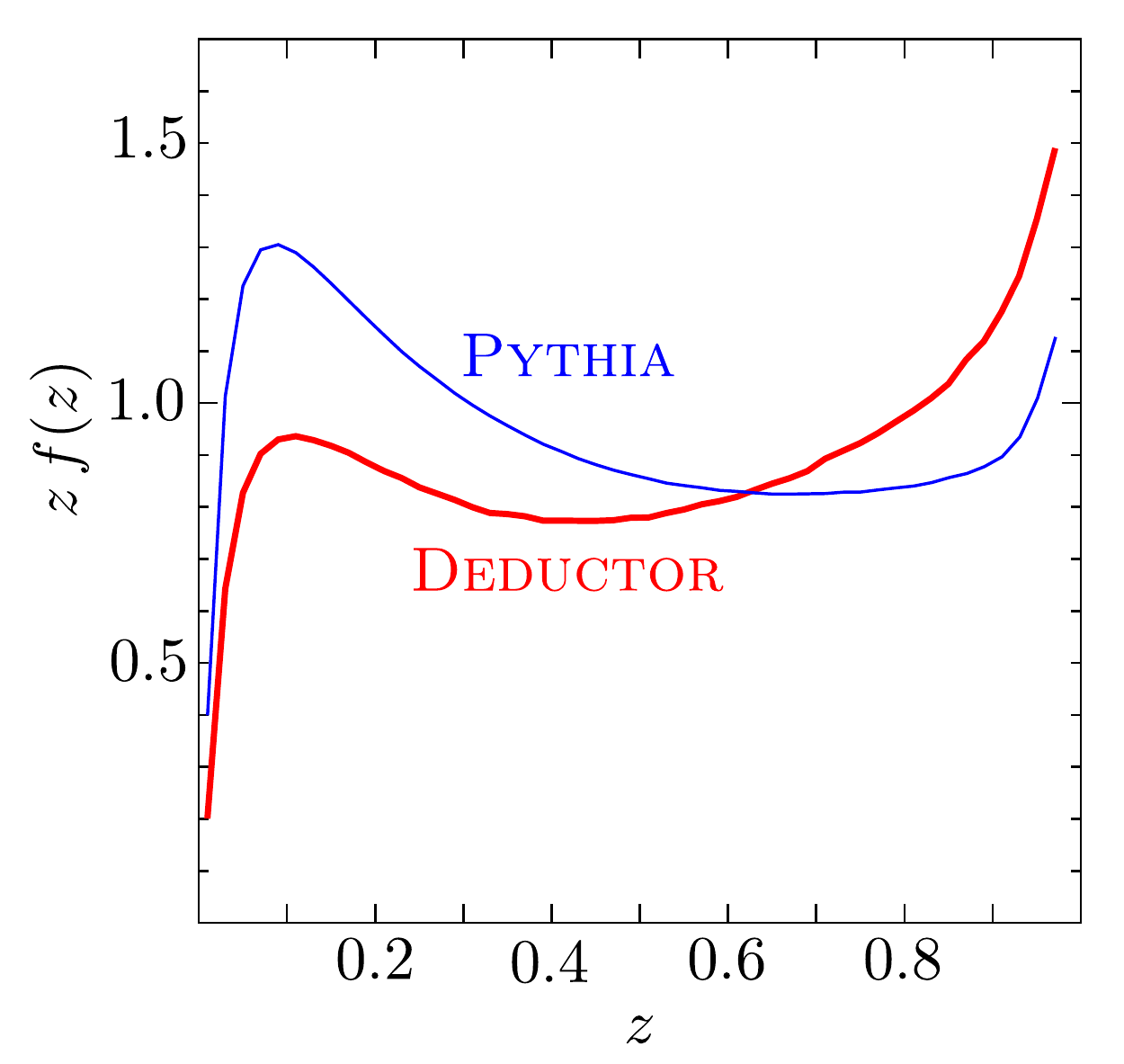}}

\caption{The $z$ distribution function of partons in a \textsc{Pythia} jet with $P_\LT > 200 \GeV$ and $|y| < 2$ compared with the same distribution for a \textsc{Deductor} jet. The jets are constructed using the $k_\LT$ algorithm with $R = 0.4$. We plot $z f(z)$.  
}
\label{fig:dNdz}
\end{figure}

Again looking at the same sample of jets as in section \ref{sec:Ninjet}, we can define a momentum fraction for each parton by $z = ({\bm p}^{\rm parton}_\LT\cdot {\bm P}_\LT^{\rm jet})/({\bm P}_\LT^{\rm jet})^2$. We can then measure the distribution of $z$ values, $f(z) = (1/N)\, dN/dz$. This distribution function obeys a momentum sum rule $\int_0^1\!dz\ z f(z) = 1$. If we measured pions instead of partons, the function $f(z)$ would be an observable, the pion decay function of a jet. It would be a non-perturbative object, but with a perturbative evolution equation. In figure~\ref{fig:dNdz}, we compare the results of \textsc{Deductor} to those of \textsc{Pythia} for $z f(z)$. There is a contribution proportional to $\delta(1-z)$ from jets that consist of one parton with $z=1$. This contribution is not seen in the plot.

We see that parton splitting above the cutoff defining the end of the shower is not quite as likely in \textsc{Deductor} as it is in \textsc{Pythia}, consistent with what we saw in section~\ref{sec:Ninjet}. Thus at the end of the shower \textsc{Deductor} has more hard partons and fewer soft partons than \textsc{Pythia}.

\subsection{Angular distribution of soft gluons between hard gluons}

In this investigation, we examine quantum interference in the emission of a soft gluon. First, we generate a sample of events with large total transverse energy $E_\LT > 400 \GeV$ (for partons with $|y| < 3.6$). In each of these events, we use the $k_\LT$ algorithm to identify jets made from partons with $|y| < 3.6$. We use a very small jet radius parameter, $R = 0.1$, so that we define the direction of the jets quite precisely. We demand that there be two jets, 1 and 2, that are fairly hard: $P_{T,1} > 50 \GeV$ and $P_{T,2} > 50 \GeV$. We further demand that the two jets be separated by an angle of about 0.5: defining $\theta^2_{12} = (y_1 - y_2)^2 + (\phi_1 - \phi_2)^2$, we demand that $0.4 < \theta_{12} < 0.6$.

\begin{figure}
\centerline{\includegraphics[width = 8 cm]{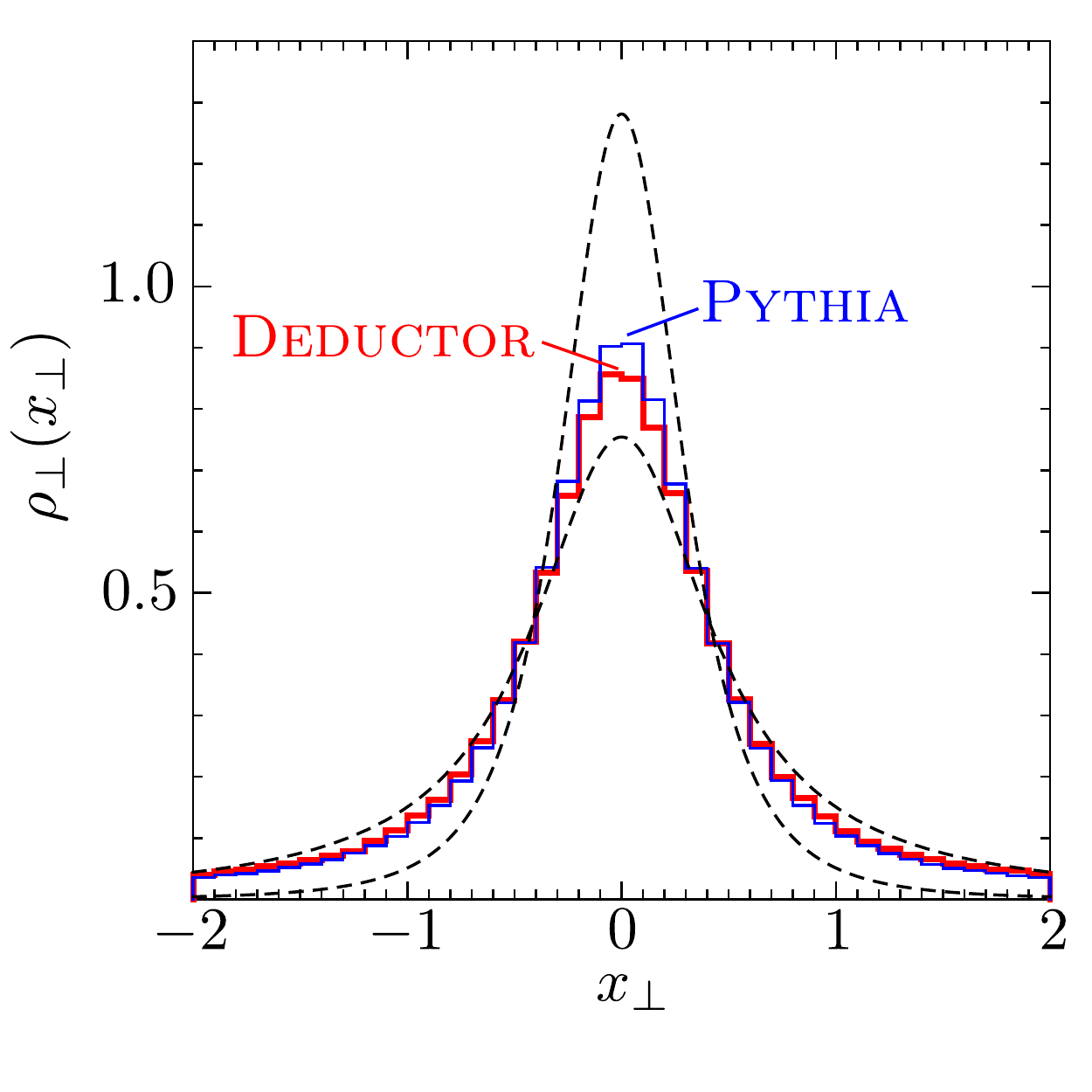}}

\caption{The distribution function, $\rho_\perp(x_\perp)$, of a soft jet as a function of the angular variable $x_\perp$ as defined in the text. The distribution from \textsc{Pythia} is compared to that from \textsc{Deductor}. The function $\rho_\perp(x_\perp; {\rm dipole})$ is displayed as the narrower dotted curve and the function $\rho_\perp(x_\perp; {\rm independent})$ is displayed as the broader dashed curve. }
\label{fig:rho}
\end{figure}

Now, suppose that we identify the jets 1 and 2 with partons and that the system including these partons emits a soft gluon. Where does the soft gluon go? To find out, we find a third jet, 3, in the event, which we imagine is identified with the soft gluon. We demand that jet 3 have $P_{T,3} > 4 \GeV$. Then $P_{T,3}$ is typically just a little bigger than this minimum value, so jet 3 is fairly soft. We are interested in the angles, $y_3$ and $\phi_3$ of the soft jet. Define angular coordinates $x_\parallel$ and $x_\perp$ by
\begin{equation}
\begin{split}
x_\parallel ={}& \frac{(y_3 - y_1)(y_2 - y_1) + (\phi_3 - \phi_1)(\phi_2 - \phi_1)}
{(y_2 - y_1)^2 + (\phi_2 - \phi_1)^2}
\;,
\\
x_\perp ={}& \frac{(\phi_3 - \phi_1)(y_2 - y_1) - (y_3 - y_1)(\phi_2 - \phi_1)}
{(y_2 - y_1)^2 + (\phi_2 - \phi_1)^2}
\;.
\end{split}
\end{equation}
Thus jet 3 is in the same direction as jet 1 when $x_\parallel = 0$ and $x_\perp = 0$, while it is in the same direction as jet 2 when $x_\parallel = 1$ and $x_\perp = 0$. The point just between the directions of the two hard jets is $x_\parallel = 1/2$ and $x_\perp = 0$. The soft jet has the same absolute angular separation from each of the two hard jets along the line $x_\parallel = 1/2$ with varying $x_\perp$.  We demand that $x_\parallel$ be near $x_\parallel = 1/2$, in a range $1/3 < x_\parallel < 2/3$. We are interested in the distribution of $x_\perp$ when $x_\parallel$ is in this range. Let $\rho_\perp(x_\perp)\, dx_\perp$ be the probability that jet 3 has the specified value of $x_\perp$ when $x_\parallel$ is in the required range and all of the other cut conditions are satisfied. We examine the range $-2 < x_\perp < 2$. Thus we normalize $\rho_\perp(x_\perp)$ to
\begin{equation}
\label{eq:xperpnorm}
\int_{-2}^{2}\! dx_\perp \,\rho_\perp(x_\perp) = 1
\;.
\end{equation}

What do we expect for $\rho_\perp(x_\perp)$? 

We may expect that $\rho_\perp(x_\perp)$ has a contribution from background jets that are not correlated with partons 1 and 2. Part of this contribution will come from initial state radiation, for instance. Thus we expect a contribution proportional to
\begin{equation}
\rho_\perp(x_\perp; {\rm background}) = \frac{1}{{\cal N}({\rm background})}
\;,
\end{equation}
where ${\cal N}({\rm background})$ is a constant (which will depend on the various parameters that go into the definitions). The factor ${\cal N}({\rm background})$ is a normalization factor. We define it and two other normalization factors below.

We also expect a contribution corresponding to independent emissions of soft gluons from either parton 1 or parton 2. This contribution will be proportional to
\begin{equation}
\rho_\perp(x_\perp; {\rm independent}) = 
\frac{1}{{\cal N}({\rm independent})}\,\frac{1}{(0.5)^2 + x_\perp^2}
\;.
\end{equation}
Here $(0.5)^2 + x_\perp^2$ is $x_\parallel^2 + x_\perp^2$ or $(1-x_\parallel)^2 + x_\perp^2$, where $x_\parallel = 0.5$ is the position of the center of the $x_\parallel$-bin.

Finally, we expect a contribution corresponding to emissions of soft gluons from the dipole consisting partons 1 and 2 (or a dipole consisting of partons moving parallel to these partons). This contribution will be proportional to
\begin{equation}
\rho_\perp(x_\perp; {\rm dipole}) = \frac{1}{{\cal N}({\rm dipole})}\,
\left[\frac{1}{(0.5)^2 + x_\perp^2}\right]^2
\;.
\end{equation}
In a partitioned dipole shower, this contribution is generated in two parts, attributed to emission from parton 1 and from parton 2. Here we put the two parts together. Note that in this contribution, emission for $|x_\perp| > 0.5$ is suppressed because of destructive quantum interference. That is, there is approximate angular ordering. However, there is no suppression for $|x_\perp| < 0.5$ because there is constructive quantum interference.

We have left normalization factors ${\cal N}$ undefined so far. We define these factors so that the distributions $\rho_\perp(x_\perp; {\rm background})$, $\rho_\perp(x_\perp; {\rm independent})$, and $\rho_\perp(x_\perp; {\rm dipole})$ are normalized on $-2 < x_\perp < 2$, as in eq.~(\ref{eq:xperpnorm}). Thus
\begin{equation}
\begin{split}
{\cal N}({\rm background}) ={}& 4
\;,
\\
{\cal N}({\rm independent}) ={}& 4 \arctan(4)
\;,
\\
{\cal N}({\rm dipole}) ={}& 32/17 + 8 \arctan(4)
\;.
\end{split}
\end{equation}

In figure~\ref{fig:rho}, we compare the results of \textsc{Deductor} to those of \textsc{Pythia} for $\rho_\perp(x_\perp)$. We see that the shapes are very similar, but that the \textsc{Pythia} curve is slightly narrower than the \textsc{Deductor} curve. We also show the functions $\rho_\perp(x_\perp; {\rm dipole})$ and $\rho_\perp(x_\perp; {\rm independent})$. To understand the results quantitatively we fit $\rho_\perp(x_\perp)$ for \textsc{Pythia} and \textsc{Deductor}  to the form
\begin{equation}
\begin{split}
\rho_\perp(x_\perp) ={}& C({\rm background})\,\rho_\perp(x_\perp; {\rm background})
\\
& + C({\rm independent})\,\rho_\perp(x_\perp; {\rm independent})
\\& +
C({\rm dipole})\,\rho_\perp(x_\perp; {\rm dipole})
\;.
\end{split}
\end{equation}
In both cases, we find that $C({\rm background})$ is small enough that we can ignore it. Then for \textsc{Pythia} we find that  $C({\rm independent}) \approx 0.68$,  $C({\rm dipole}) \approx 0.32$. For \textsc{Deductor},  $C({\rm independent}) \approx 0.78$,  $C({\rm dipole}) \approx 0.22$. We judge that the difference between \textsc{Pythia} and \textsc{Deductor} is not very important but that it is important that one can see the quantum interference effect that is built into both these programs.

\subsection{Transverse momentum distribution of vector bosons}
\label{sec:DrellYan}

We examine next the production $e^+ e^-$ pairs produced from a Z-boson or photon in a proton-proton collision with $\sqrt s = 8 \TeV$. We are interested in the transverse momentum, $p_\LT$, of the $e^+ e^-$ pair. We demand that the mass of the $e^+ e^-$ pair be large, $M(e^+ e^-) > 400 \GeV$, and that its rapidity be in the central region, $|y(e^+ e^-)| < 2$, and we look at the region of small and moderate transverse momentum, $0 < p_\LT < 100 \GeV$.  We measure the distribution $\rho_\LZ(p_\LT) = (1/\sigma)\, d\sigma/dp_\LT$ for this event sample. The normalization is $\int_0^{100 \GeV}\!dp_\LT\ \rho_\LZ(p_\LT) = 1$.

\begin{figure}
\centerline{\includegraphics[width = 8 cm]{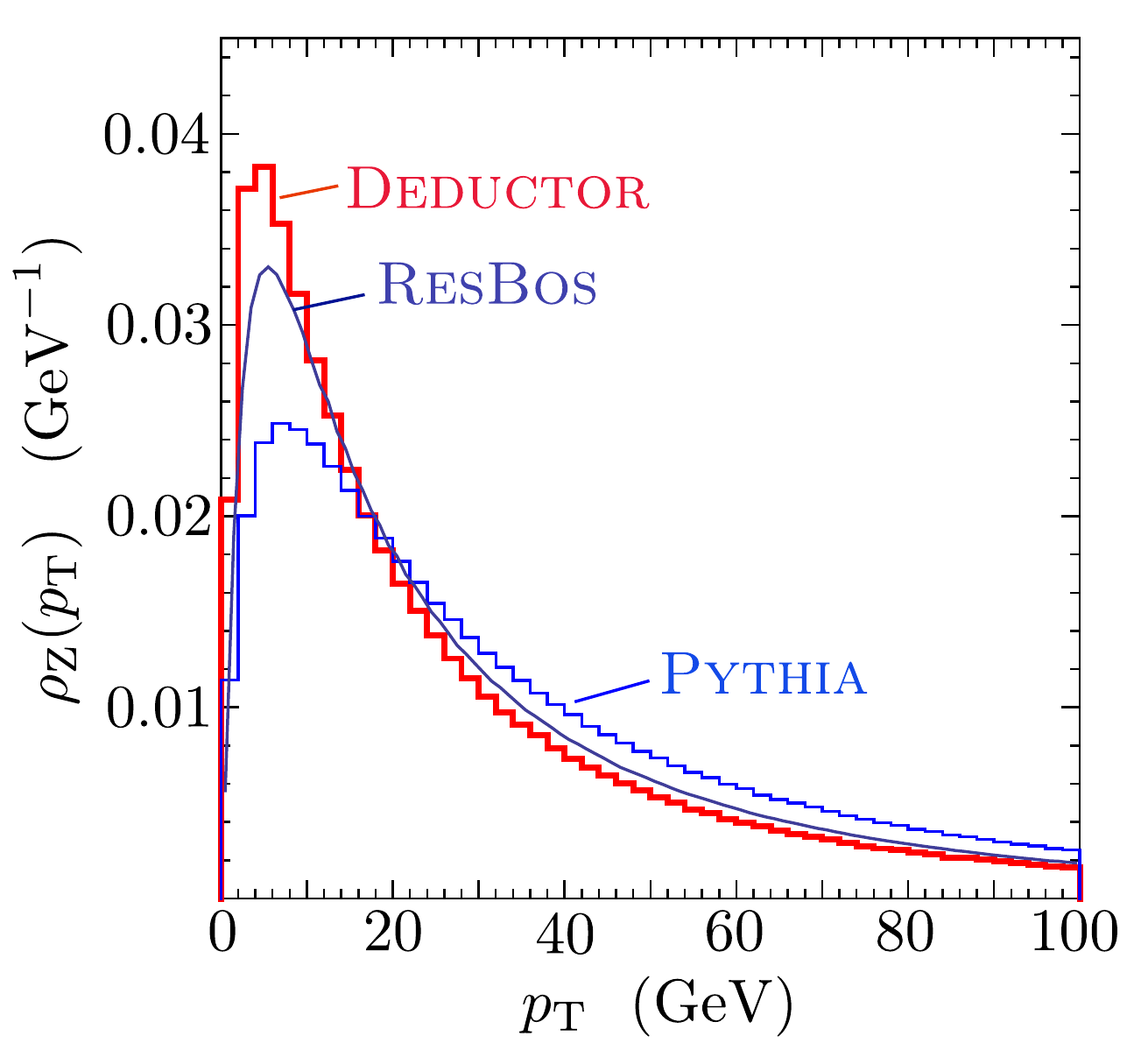}}

\caption{The transverse momentum distribution $\rho_\LZ(p_\LT)$ of $e^+ e^-$ pairs produced in the Drell-Yan process. We compare the distribution from \textsc{Deductor} to that \textsc{Pythia} and also to the distribution from \textsc{ResBos}, which sums logs of $p_\LT/M(e^+e^-)$. See the text for details. In the left hand portion of the figure, the top curve is \textsc{Deductor}, the middle curve is \textsc{Resbos}, and the lower curve is \textsc{Pythia}}
\label{fig:ZpT}
\end{figure}

We compare the result from \textsc{Deductor} to that from \textsc{Pythia} in figure \ref{fig:ZpT}. We see that the curve from \textsc{Deductor} is somewhat narrower than that from \textsc{Pythia}. There are non-perturbative effects that are missing in \textsc{Deductor} and have been turned off in \textsc{Pythia}. There are also effects from the choice of how the perturbative shower is ended. These effects can have the effect of changing the width of the distributions and are especially important for $p_\LT < 10 \GeV$. Thus it seems plausible that non-perturbative and shower-end effects account for the difference between the \textsc{Deductor} and \textsc{Pythia} curves.
 
There is a theoretical result \cite{CSSpT} for the structure of $\rho_\LZ(p_\LT)$ after summing logs of $p_\LT/M$. This result, including a fit to nonperturbative effects that smear the distribution slightly, is available in the computer program \textsc{ResBos}, by C.~Balazs, P.~Nadolsky, and C.P.~Yuan \cite{ResBos1, ResBos4, ResBos20}. We also show a result from \textsc{ResBos} in figure \ref{fig:ZpT}. Our \textsc{ResBos} results are for a virtual $Z$-boson rather than a linear combination of a virtual $Z$-boson and a virtual photon as simulated in \textsc{Deductor} and \textsc{Pythia}. In the region that we study, this difference should affect the normalization of the cross section but its effect on the shape should be small.

A dipole shower, as in \textsc{Deductor}, approximately sums the logs of $P_\LT/M$ \cite{NSDrellYan}.\footnote{As noted in section~\ref{sec:settings}, in \textsc{Deductor}, we set the $\as$ scale for an initial state splitting to $\lambda_\LR k_\LT^2$ where $k_\LT$ is the transverse momentum in the splitting and where $\lambda_\LR$ is given in eq.~(\ref{eq:lambdaR}). This aids in generating the proper subleading logarithms \cite{NSDrellYan, CataniMCsummation} in the vector boson transverse momentum distribution.} Thus we expect \textsc{Deductor} and \textsc{ResBos} results for $\rho_\LZ(p_\LT)$ to agree approximately. It seems plausible that adjusting how the shower ends and including nonperturbative effects can broaden the \textsc{Deductor} result slightly to bring it into better agreement with \textsc{ResBos}.

\subsection{Transverse momentum distribution of associated b-quarks}

Initial state b-quarks and c-quarks are have non-zero masses in \textsc{Deductor}
but are treated as massless in \textsc{Pythia}. Does this make a difference? To investigate this question, we simulate the production of an $e^+ e^-$ pair in the Drell-Yan process, as in the previous subsection. However we demand that the $\LZ$ or $\gamma$ be produced by a collision of a $\Lb$-quark and a $\bar \Lb$-quark, where the b-quark comes from the proton with positive $p^3$. Of course, a $\LZ$ or $\gamma$ can be created by annihilation of any flavor of quark with its corresponding antiquark, but we examine only the b-quark process. Consider the b-quark that annihilates to make the $\LZ$ or $\gamma$. In the backwards evolution of the initial state, this b-quark eventually becomes a gluon, with the emission of a $\bar \Lb$ quark.\footnote{Very rarely, our requirement that $k_\LT^2 > 1\ \GeV$ for allowed shower splittings does not allow a b-quark to become a gluon. In this case we consider the $\bar \Lb$ to be part of the beam with zero transverse momentum and infinite rapidity.} Where does the $\bar \Lb$ quark go?

\begin{figure}
\centerline{
\includegraphics[width = 7.5 cm]{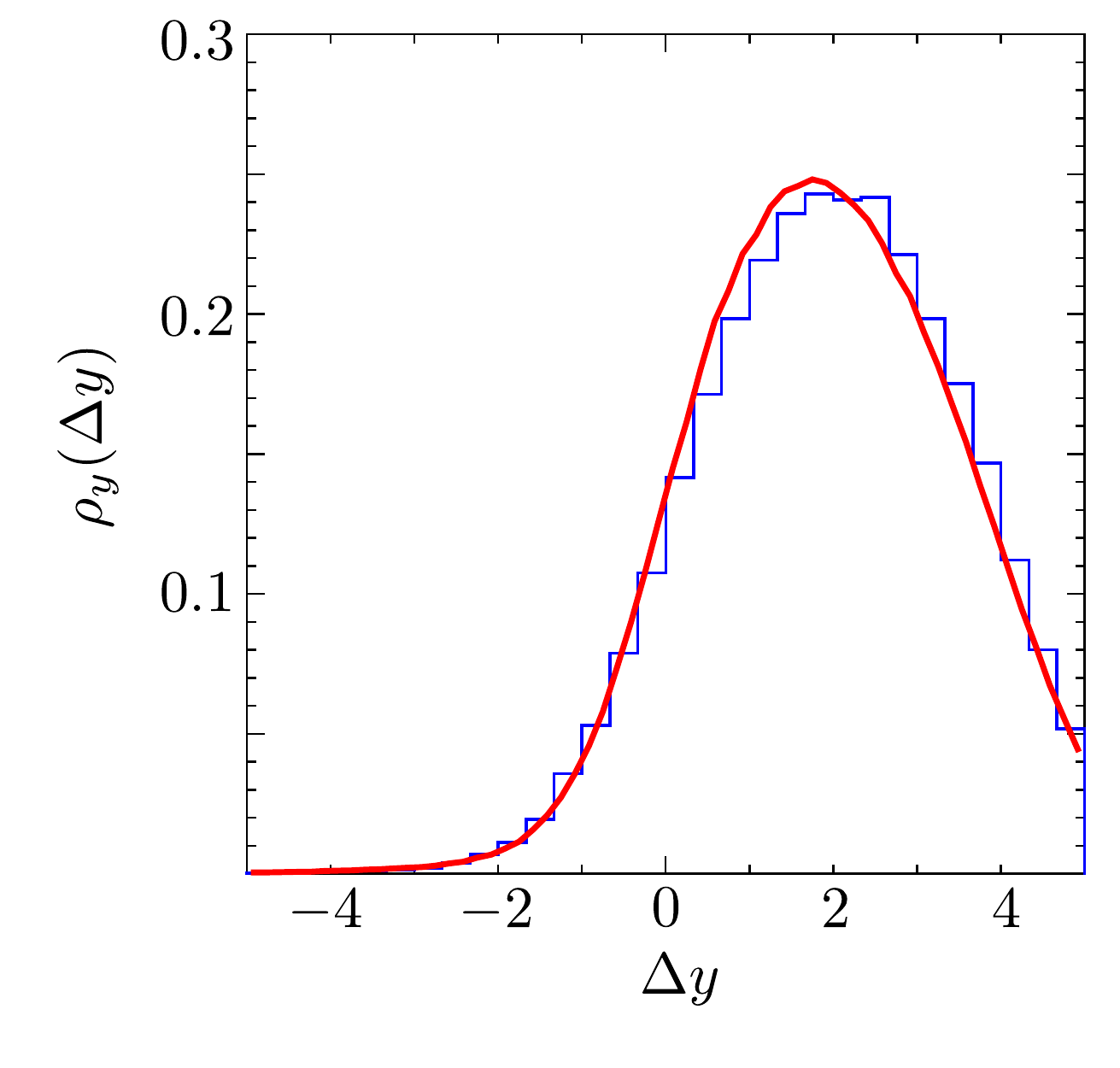}
\includegraphics[width = 7.5 cm]{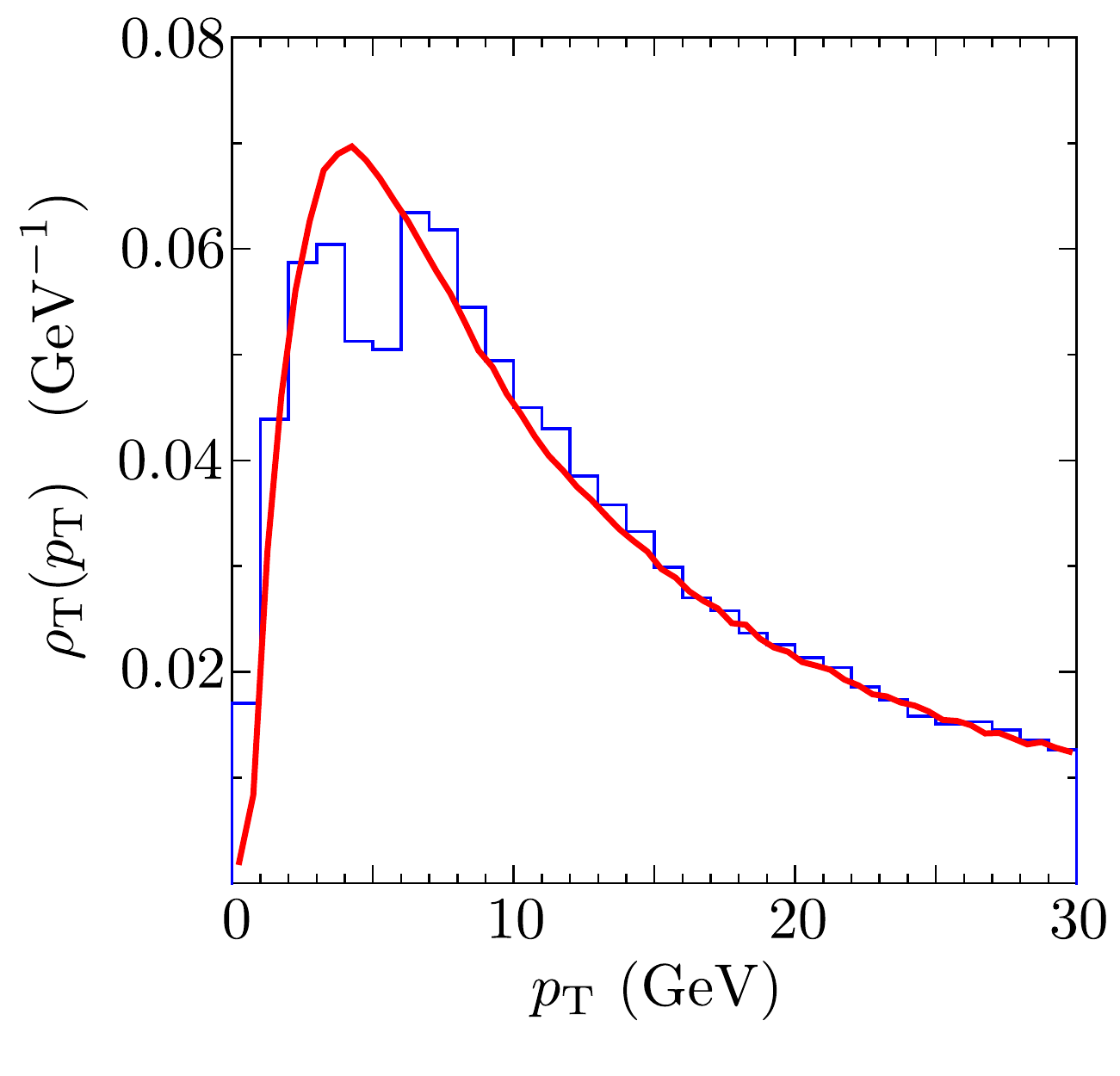}
}

\caption{The distribution of $\bar \Lb$-quarks in events with the hard process $\Lb + \bar \Lb \to \LZ/\gamma + X \to e^++  e^- + X$ with $M(e^+ e^-) > 400 \GeV$ and $|y(e^+ e^-)| < 2$. The left hand plot compares the distribution in rapidity $\Delta y = y(\bar \Lb) - y(e^+e^-)$ from \textsc{Deductor} (the curve) and \textsc{Pythia} (the histogram). The right hand plot compares the distributions in transverse momentum $p_\LT(\bar \Lb)$. Again, the curve is from \textsc{Deductor} and the histogram is from \textsc{Pythia}.}
\label{fig:bbardist}
\end{figure}

We define $\rho_{Y}(\Delta y)\, d\Delta y$ to be the probability for the $\bar \Lb$ quark to have rapidity $\Delta y$ relative to the rapidity of the $e^+ e^-$ pair, $\Delta y = y(\bar \Lb) - y(e^+e^-)$, normalized to $\int_{-5}^5\!d\Delta y \, \rho_{Y}(\Delta y) = 1$. In the left hand plot of figure~\ref{fig:bbardist}, we compare the result for $\rho_{Y}(\Delta y)$ from \textsc{Deductor} to that from \textsc{Pythia}. We see that there is hardly any difference.

We define $\rho_{\LT}(p_\LT)\, dp_\LT$ to be the probability for the $\bar \Lb$ quark to have transverse momentum $p_\LT$, normalized to $\int_{0}^{30 \GeV}\!dp_\LT\, \rho_{\LT}(p_\LT) = 1$.  In the right hand plot, we compare the result for $\rho_{\LT}(p_\LT)$ from \textsc{Deductor} to that from \textsc{Pythia}. There is hardly any difference except near $p_T$ equal the the b-quark mass $m(\Lb)$, which is 4.75 GeV in \textsc{Deductor}. Near $p_\LT = m(\Lb)$, the \textsc{Deductor} curve has a peak. This seems sensible according to the kinematics of the process $\Lg \to \Lb + \bar \Lb$. Near $p_\LT = m(\Lb)$, the \textsc{Pythia} histogram has a dip. Presumably this is because initial state quarks are treated as massless in \textsc{Pythia}. These results are before transverse momentum smearing of the incoming partons and before hadronization. We expect that after these effects, the \textsc{Pythia} dip will be washed out.

\section{Conclusions}
\label{sec:conclusions}

We have presented some first results from a new parton shower Monte Carlo event generator, \textsc{Deductor}, which generates events from a hard scattering process and both initial state and final state parton showers. Hadronization and an underlying event are not included in this initial version. Our main purpose in creating this event generator was to investigate the effects of parton color and spin. However, in this paper, we confine our investigation to the leading color, spin averaged approximation. The code implements also the LC+ approximation \cite{NScolor}, which we will compare to the leading color approximation in a future work. There is a straightforward algorithm \cite{NSspin} for improving the treatment of spin, but this algorithm is not yet implemented.

There are some new features in \textsc{Deductor} compared to other parton shower event generators. Shower splittings are generated according to a virtuality based ordering variable $\Lambda^2$ defined in eq.~(\ref{eq:showertime}) instead of transverse momentum $k_\LT$. This is discussed in a companion paper \cite{NSShowerTime}. \textsc{Deductor} does not follow the standard practice of setting the masses of initial state b and c quarks to zero. This requires new evolution kernels for the parton distribution functions. This is discussed in a companion paper \cite{NSpartons}.

We have compared results from \textsc{Deductor} to results from \textsc{Pythia} for a number of distributions that illustrate the workings of a parton shower. We find that there are some differences but that they are not large. As seen in figure~\ref{fig:rho}, both programs exhibit the effect of quantum interference between soft gluon emissions from a color dipole. With the shower-end settings that we used, \textsc{Deductor} has somewhat fewer splittings than \textsc{Pythia}. Jet cross sections are comparable, as are the transverse momentum distributions of Drell-Yan lepton pairs. By looking in the right place, one can observe differences that result from the different treatment of parton mass for initial state b-quarks.

The comparisons that we have generated suggest that \textsc{Deductor} is working sensibly. It remains to investigate the effects of including better approximations for color and spin.

\acknowledgments{ 
This work was supported in part by the United States Department of Energy and by the Helmholtz Alliance ``Physics at the Terascale." We thank Voica Radescu of the \textsc{HeraFitter} group for providing the parton distribution functions that we use. We thank Pavel\ Nadolsky and C.P.\ Yuan for help with \textsc{ResBos}. We thank Judith Katzy for helpful conversations.}


\end{document}